# If a tree casts a shadow is it telling the time?

Russ Abbott[1]

[1]Department of Computer Science, California State University, Los Angeles, Ca, USA
Russ.Abbott@GMail.com

**Abstract.** Physical processes are computations only when we use them to externalize thought. Computation is the performance of one or more fixed processes within a contingent environment. We reformulate the Church-Turing thesis so that it applies to programs rather than to computability. When suitably formulated agent-based computing in an open, multi-scalar environment represents the current consensus view of how we interact with the world. But we don't know how to formulate multi-scalar environments.

**Keywords:** agents, agent-based, agent-based computation, Church-Turing thesis, Church's thesis, computing, computation, environment ideas, interaction, interactive computation, models, multi-scalar environment, thought, thought tools, unconventional computation.

## 1 Introduction

In the preface to the first edition of the *International Journal of Unconventional Computation*, the editorial board [1] welcomed papers in "information processing based on physics, chemistry and biology." But the Board left undefined what it means to say (a) that a physical, chemical, or biological system is doing "information processing" or (b) that information processing is "based on physics, chemistry, or biology." In this paper we explore these issues by focusing on these questions.

- *What is computation?*
- *How can computation be distinguished from other natural processes?*
- *What is the relationship between ideas and computations?*
- *What is the relationship between a computational process and the environment within which it occurs?*

Our conclusions will be that physical processes are considered computation when we treat them as externalized thought and that computation itself involves the playing out of fixed processes against a contingent environment. We re-interpret the Church-Turing thesis: for a thought to be rigorous it should, at least in principle, be expressible as a computer program. A corollary is that we agree with Wegner [2] that the agent-based model of computation is the right way to think about interaction with an environment. But since we don't understand how to model multi-scalar environments we are not in a position to model our understanding of how we interact with nature.

## 1.1 Is Google reading my email?

That's the first question in the Google Gmail help center [3]. This question arises because Gmail places ads next to email messages, and the selection of ads is based on the contents of the messages. Google's answer to this question has varied over time. On March 13, 2006, the posted answer was as follows.

> Google computers scan the text of Gmail messages in order to filter spam and detect viruses, just as all major webmail services do. Google also uses this scanning technology to deliver targeted text ads and other related information. *The process is completely automated and involves no humans.* [Emphasis added.]

In other words, Google's computers are reading your email—but no human beings are. That most people find this reassuring illustrates the intuition that *it's what goes on in the mind of a human being that matters to us.*

One might object that if a computer is reading one's email (and storing its contents in a database), a person might read it later. That's quite true, and the fact that only Google computers (and not Google employees) are reading one's email when selecting ads does not guarantee one's privacy. But if no person *ever* reads one's email, then most people will not feel that their privacy has been violated.

After all, all email is read by a number of computers as it passes from sender to receiver. No one has ever worried about that. The moment of violation occurs when some living human being becomes consciously aware of one's personal information.

But, one might argue, the kind of reading that occurs when a computer transmits a message along a communication channel is different from the kind of reading that occurs when a Google computer determines which ads to place next to a message. The former treats messages as character strings; no meaning is extracted. The later extracts (or attempts to extract) meaning so that related ads can be displayed.

This raises the question of what we understand by the term *meaning*. That's clearly a larger topic than we can settle here, but our short answer is that our intuitive sense of *meaning* has something to do with an idea or thought forming in a mind.[1] At this stage in the development of technology, most people don't believe it makes sense to say that an idea has formed in the mind of a computer—or that a computer has a mind at all. We may speak informally and say something like "the computer is doing this because it thinks that." But when we say these sorts of things, we are speaking metaphorically. Until we start to think of computers as having minds that have subjective experience, minds in which ideas can form—then most people will feel comfortable with Google's reply that its computers, but no human beings, are reading one's email.

## 1.2 To come

Section 2 continues the discussion of thoughts and introduces the notion of thought tools, for which it provides a brief history. Section 3 considers how *computation* might be defined. Section 4 discusses the agent-based computing paradigm as more than just an approach to programming and modeling but as common to many of the ways we think about both thinking and our interaction with nature.

---

[1] This clearly is different from the formal semantics sense in which *meaning* refers to a mapping from an expression to a model.

## 2  Thinking and thought tools

If a tree grows in a forest, but no one counts its rings is it counting years? Is it performing an unconventional computation? If a tree grows in a forest but no one knows it's there, is it instantiating the idea of a tree? These questions have the same sort of answers as does Bishop Berkeley's famous question: if a tree falls in a forest with no one around to hear it, does it make a sound?

Berkeley's question is not as difficult as it seems. Our answer, which is different from Berkeley's,[2] is that one must distinguish between physical events and subjective experience. If a tree falls in a forest, it generates (what we call) *sound waves* whether someone is there to hear them or not. But if no one is there to hear the sound, if no being has a subjective experience of the sound, then no (subjective) sound is heard.

The same holds for ideas. Like the subjective experience of a sound, the *idea* of a tree exists only as a subjective experience. If no one has that subjective experience, then a tree without anyone knowing about it will not be instantiating the idea of a tree. Even if one grants that the idea of a tree is exactly the right way to describe that particular aspect of nature, that idea exists only as an idea, and it exists only in the mind of someone who is thinking it. Ideas exist only within the realm of mental events, i.e., as subjective experience. In saying this we are taking an explicitly anti-Platonist stance: there is no realm outside the mind in which ideas exist on their own.[3]

This is not intended as mystical or profound—just a statement of a brute fact: an idea is something that occurs only in someone's mind. The ideas in this paper exist only in the mind of the author and the minds of the readers as the author and readers are thinking them. These ideas don't exist on the paper or on the computer screens on which these words appear. They don't exist in the computer memory in which these words are stored. Just as the moment at which an invasion of privacy occurs is when some being-with-a-mind learns something personal about us, an idea exists only when someone is thinking it.[4] We go to such lengths to make this point because our position is that computations, like ideas, are also mental events, but mental events that we have externalized in a way that allow us to use physical processes to perform them.

When a tree grows rings, it just grows rings. But when we use that tree-ring growth as a way to count years, i.e., to help us work with ideas such as the idea of a year, then we can say that the tree has performed a computation—an unconventional one.

---

[2] Berkeley's answer is that it makes a sound because God, who is always everywhere, hears it.

[3] We are not taking a stand on nominalism vs. realism. Although we believe that our (human) ideas about how nature should be described are not arbitrary and that entities other than the elementary particles exist (see Abbott [4]), that is not at issue here.

[4] This position requires some care in formulation. If an idea exists only when someone is thinking it, what does it mean to say that two people have or had "the same" idea? We believe that these issues can be worked out.

When a computer runs is it computing? Our answer is the same. A computer is computing only when it is understood to be performing some externalized mental activity. Otherwise, it's just an arena within which electrons are moving about.

**2.1  A brief history of the internalization and then the externalization of thought**

One may trace one thread through the history of thought as the internalization and then the externalization of thought. Initially we looked outward for answers to questions about how to make sense of the world. Not knowing what else to do, we looked to sources of what we hoped were authority: priests, oracles, prophets, sacred writings, divinities, etc., to tell us what thoughts to install in our minds.[5]

We often fought with each other about whose sources of knowledge were right. In a recent op-ed piece [5] Lorenzo Albacete, a Roman Catholic priest, articulated the position of those who fear the use of religion as a source of knowledge.

> For [nonbelievers], what makes Christianity potentially dangerous [is not its other-worldliness but] its insistence that faith is … the source of knowledge about this world.

As Albacete later notes, by the time of the Roman Empire, the use of religion as a source of ideas about how nature works had been discarded by enlightened thinkers. Greek and Roman philosophers believed that they themselves could be a source of knowledge about the world. The step from looking for external sources of knowledge to supposing that perhaps we can figure it out for ourselves is what we are referring to as the internalization of thought—attributing to oneself the power to produce thoughts of value and rejecting the notion that thoughts must originate externally to be valid.

**2.2  Externalizing thought and tools to work with it**

The history of early computing may be traced along three paths. Each path traces devices that help us think about a particular (and fundamental) subject area: time, counting (arithmetic), and space (geometry).

**2.3  Time computers**

We used natural processes to help us express our ideas about time—the daily, monthly, and yearly cycles of the earth, moon, and sun. Not to beat this point into the ground, *day, month,* and *year* are ideas. As ideas, they exist only in the mind—no matter how accurate or true they are as descriptions of nature.

---

[5] One wonders what priests, oracles, prophets, and other human authorities believed about how the ideas they transmitted arrived in their own minds. Perhaps they believed that the ideas had been implanted in their minds as a result of their special status or as a result of some special words or rituals that they performed. Perhaps they were just transmitting ideas that had been transmitted to them. Presumably they didn't believe that they themselves made up these ideas. Most likely they didn't ask themselves this question.

The first (analog[6]) time computers were the actual processes that corresponded to our thoughts. The rising and setting of the sun were the physical events that we used to keep track of the mental events: the *start* and *end* of a *day*. Similarly for the moon. Yearly events such as river floodings and the comings and goings of the seasons helped us keep track of the mental event: the yearly cycle.

It didn't take us long to invent more sophisticated analog computers. The sundial, for example, is an analog computing device. The position of the sun's shadow is an analog for the mental event *time-of-day* which corresponds to the physical relationships between the relative positions of the sun and the earth.

It is worth noting that with the sundial we started to arrange physical materials to help us track our thoughts. In building sundials we set up shadow casters, which in conjunction with the sun and markings on the surface on which the shadow is cast, helped us track (our ideas about) the passing of the day. Presumably this was not a very significant step from using existing shadow-casting objects, e.g., trees, for the same purpose. Hence our title: if a tree casts a shadow, is it telling the time?

**2.4 Number computers**

Apparently we started to count quite early. Bones with notches carved into them appeared in western Europe 20,000 to 30,000 years ago. There is evidence of the use of a tally system—groups of five notches separated from each other. With tally systems not only did we mark physical materials to help us keep track of numbers (which are also mental events), we also invented ways to make counting easier by the way in which we arranged these markers, i.e., in groups. Soon we invented the abacus.

With these primitive computers we separated the computational process from its dependency on natural processes. Sundials and astronomical masonry depend on the sun and the stars. Counting depends on nothing other than human activity. Once we invented computational devices that were independent of non-human physical processes it was a short step to written notation. By approximately 3,000 BC cuneiform writing on clay tablets using positional notation was known in Babylonia.

**2.5 Space computers**

Besides time and numbers, the Pythagoreans in Greece and Euclid in Egypt developed ways to think about space. We know that early geometers thought about construction issues. The straight edge and compass were their (human-powered) thought tools. They used them to externalize, to create representations of, and to manipulate the ideas of straight lines and circles.

**2.6 Is it reasonable to call abaci and geometers' tools computers?**

Even though abaci and geometers' tools are completely independent of non-human physical processes, i.e., they are entirely dependent on human activity to make them

---

[6] An analog computer is so called because can be understood as analogous to something else.

"run," we feel justified in calling them computers because they are used according to mechanical rules. Even though the source of energy for an abacus is the user, the abacus user follows strict rules—rules which could be automated.

### 2.7 Thought tools for symbol manipulation

Beyond time, numbers, and space, we have also built thought tools to represent symbolic thoughts and relationships. Sowa [6] describes the Tree of Porphyry.

> The oldest known semantic network was drawn in the 3rd century AD by the Greek philosopher Porphyry in his commentary on Aristotle's categories. Porphyry used it to illustrate Aristotle's method of defining categories by specifying the *genus* or general type and the *differentiae* that distinguish different subtypes of the same supertype.

Another attempt to externalize symbolic thought has been credited to Ramon Lull in the late $13^{th}$ century. Smart [7] describes it as follows.

> Ramon Lull's logic machine consisted of a stack of concentric disks mounted on an axis where they could rotate independently. The disks, made of card stock, wood, or metal, were progressively larger from top to bottom. As many as 16 words or symbols were visible on each disk. By rotating the disks, random statements were generated from the alignment of words. Lull's most ambitious device held 14 disks.
>
> The idea for the machine came to Lull in a mystical vision that appeared to him after a period of fasting and contemplation. It was not unusual in that day … scientific advances to be attributed to divine inspiration. He thought of his wheels as divine, and his goal was to use them to prove the truth of the Bible. …
>
> In "Gulliver's Travels," Swift satirizes the machine without naming Lull. In the story, a professor shows Gulliver a huge contraption that generates random sequences of words. Whenever any three or four adjacent words made sense together, they were written down. The professor told Gulliver the machine would let the most ignorant person effortlessly write books in philosophy, poetry, law, mathematics, and theology.

This may be the first use of non-determinism in computing.

Soon thereafter William of Ockham discovered the foundations of what were to become De Morgan's laws of logic. More specifically, from Sowa [8]:

> (Ockham, 1323) showed how to determine the truth value of compound propositions in terms of the truth or falsity of their components and to determine the validity of rules of inference … in terms of the truth of their antecedents and consequents.

### 2.8 Thought tools and the scientific process

Clocks, abaci, straight-edges, hierarchies, non-determinism, laws of logic, and other thought tools differ in kind from microscopes, telescopes, and other scientific instruments of observation. The former are intended to allow us to externalize and manipulate our thoughts. The latter allow us to investigate nature—to see what's out there and perhaps to see things that will require new ideas to understand them. Thought tools are constructive; instruments of scientific observation are reductive.

After having convinced ourselves that we are capable of generating our own ideas, an important next step was to realize the necessity of testing our ideas against nature. Simply coming up with an idea is not enough. It's important both to externalize it as a way to work with it and to test it by looking at nature though it. Thus science consists fundamentally of three kinds of activity. (We are not arguing that these activates occur sequentially.)

1. **Observation.** Uncovering new facts about nature.
2. **Reduction/invention.** Reverse engineering nature to figure out how it may have harnessed understood principles to produce the observed facts. Although reverse engineering, i.e., reductionism,[7] sounds unglamorous, it is fundamental. Determining that our genome is encoded as a double helix was reverse engineering. The invention aspect of reverse engineering is to imagine the design that nature uses to achieve an end.
3. **Creation/invention.** Establishing new fundamental principles and then using them as the basis of the reverse engineering process. This occurs only in fundamental physics.

Scientific instruments help us with (1). Thought tools help with (2) and (3).

**2.9 The state of the art of thought externalization**

Every computer application is a thought tool. The thoughts that are being manipulated are the thoughts that are represented by the conceptual model implemented by the application. More importantly every programming language is a thought tool. Programming languages allows us to externalize in the form of computer programs our thoughts about symbolic behaviors. Since one writes computer applications in programming languages, a programming language is a thought tool for building thought tools, i.e., a thought tool for externalizing thought.

It is important to realize that a programming language is itself a computer application. As a computer application, it implements a conceptual model; it allows its users to express their thoughts in certain limited ways, namely in terms of the constructs defined by the programming language. Since all modern programming languages are conceptually extensible—libraries allow us to extend the conceptual constructs defined by the language—programming languages allow us to define concepts, which we can then use to build other concepts. As such they are very powerful thought tools.

We are still learning how to use the power of computers to externalize thought. In one way or another, much of software-related research is about developing more powerful, more specialized, faster, easier to use, or more abstract thought tools. The more we learn about externalizing our thoughts the higher we ascend the mountain of abstraction and the broader the vistas we see. It seems appropriate that one current avenue of research, language for building models (UML and SysML) and languages for describing ontologies (the OWL Web Ontology Language), continue a tradition that dates back to Porphyry—and before.

---

[7] Reductionism has recently received a lot of bad press. As explicated here, the reductionist impulse often leads to the development of important new ideas.

## 3   Defining computation

In this section we turn to the question of how to define *computation*. It is surprisingly difficult to find a well considered definition. The one offered by Eliasmith [9] appears to be the most carefully thought out. Here is his definition and his commentary.

> *Computation*. A series of rule governed state transitions whose rules can be altered. There are numerous competing definitions of *computation*. Along with the initial definition provided here, the following three definitions are often encountered:
> 1. Rule governed state transitions
> 2. Discrete rule governed state transitions
> 3. Rule governed state transitions between interpretable states
>
> The difficulties with these definitions can be summarized as follows:
> a) The first admits all physical systems into the class of computational systems, making the definition somewhat vacuous.
> b) The second excludes all forms of analog computation, perhaps including the sorts of processing taking place in the brain.
> c) The third necessitates accepting all computational systems as representational systems. In other words, there is no computation without representation on this definition.

Contrary to Eliasmith we suggest the following.

a) The notion of alterable rules is not well defined, and hence all physical systems *are* potentially computational systems.

b) But, it is exactly the fact of interpretability that makes a physical process into a computation. (Eliasmith doesn't explain why he rejects the notion that computation requires interpretation.)

Eliasmith requires that the rules governing some identified state transitions must be alterable in order to distinguish a computation from a naturally occurring process—which presumably follows rules that can't be altered. But all computing that takes place in the physical world is based on physical processes. If we set aside the probabilistic nature of quantum physics, and if we suppose that physical processes operate according to unalterable rules, it's not clear what it means to say that it must be possible to alter a set of rules.

This is not just being difficult. Certainly we all know what it means to say that one program is different from another—that "the rules" which govern a computation, may be altered. But the question we wish to raise is how can one distinguish the altering of a program from the altering of any other contingent element in an environment?

It is the particular program that is loaded into a computer's memory that distinguishes the situation in which one program is being executed from that in which some other program is executing. But a computer's memory is the environment within which the computer's cpu (or some virtual machine) finds itself, and a loaded program defines the state of that environment. The cpu (or the virtual machine) is (let's presume) fixed in the same way that the laws of nature are fixed. But depending on the environment within which it finds itself—i.e., the program it finds in its environment—the cpu operates differently, i.e., it performs a different computation.

This same sort of analysis may be applied to virtually any natural process. When we put objects on a balance scale, the scale's behavior will depend on the objects loaded, i.e., on the environmental contingencies.[8] In both the case of programs loaded into a computer and objects put in the pans of a balance scale, we (the user) determine the environment within which some fixed process (i.e., the rules) proceeds.

This brings us back to our original perspective. A process in nature may be considered a computation only when we use it as a way to work with externalized thought. A physical or otherwise established process—be it the operation of a balance scale, a cpu, the Game of Life, or the sun in motion with respect to trees and the ground—is just what it is, a fixed process.[9] But for almost all processes,[10] whether we create them or they arise naturally, how the process proceeds depends on environmental contingencies. When we control (or interpret) the contingencies so that we can use the resulting process to work with our own thoughts, then the process may be considered a computation. This is the case whether we control the contingencies by loading a program into a computer, by placing objects on a balance scale, by establishing initial conditions for the Game of Life, or by giving meaning to shadows cast by trees.

Consequently we agree with Eliasmith that it must be possible to alter a process for it to be considered a computation, but we would express that condition in other words. For a process to be considered a computation there must be something contingent about the environment within which it operates which both determines how it proceeds and determines how we interpret the result.

In other words, we can always separate a computational process into its fixed part and its contingent or alterable part. The fixed part may be some concrete instances of the playing out of the laws of nature—in which case the contingent environment is the context within which that playing out occurs. Or it may be the operation of a cpu—in which case the contingent environment is the memory which contains the program that is being executed. Or it may be the operation of a program that a cpu is executing—in which case the contingent environment is the input to that program. A computation occurs when we alter the contingencies in the environment of an fixed process as a way to work with our thoughts.

This perspective contrasts traditional (theoretical) computation with real-world computation. Traditionally, one thinks of a computation as a contingent process—one defined in a programming language. Like a Turing Machine, it runs for free. Real-world computations result from non-contingent processes, have energy requirements, and operate in contingent environments.

---

[8] When a balance scale compares two objects and returns an "output" (selected from *left-is-heavier*, *equal-weights*, and *right-is-heavier*), is it performing a computation? It is if we are using it for this purpose. It isn't if we are using it as a designer setting for flower pots.

[9] Of course many processes—such as the operation of a cpu and the operation of a balance scale—are what they are because we built them to be that way—because we anticipated using contingencies that we could control in their environment to help us think.

[10] Some quantum processes may occur on their own without regard to their environment—although even they are environmentally constrained by the Pauli exclusion principle..

### 3.1 Non-algorithmic computing

A corollary of the preceding is that all computation performed by real-world processes are environmentally driven. Computing involves configuring environmental contingencies, i.e., setting up an environment within which a process (or multiple processes) will play itself (or themselves) out. We refer to this as *non-algorithmic computing*. The focus is on how an environment will shape a process, not on the algorithm the shaped process will perform. A Game of Life glider has no algorithm.

It may seem ironic that what we think of as conventional computation on real-world computers is a constrained form of unconventional computation. We are attracted to it because its single threaded linearity makes it easy to manage. But nature is not linear. Any computer engineer will confirm how much work it takes to shape what really goes on in nature into a von Neumann computer.

Perhaps even more ironically, we then turn around and use conventional single-threaded computers to simulate nonlinear unconventional computation. One might say that a goal of this conference is to eliminate the von Neumann middle man—to find ways to compute, i.e., to externalize our thoughts, by mapping them more directly onto the forces of nature operating in constrained environments.

### 3.1 Turing *Machines* vs. Turing *computability*

Why can't we look to Turing Machines (and their equivalents) for a definition of computation that does not rely on the notion of thought externalization? Turing Machines, recursive functions, and equivalent models rely on the notions of symbols and symbol manipulation, which are fundamentally mental constructs. Eliasmith's definition doesn't—although his definition does depend on the notion of rule-governed state transitions, which appears difficult to define non-symbolically. The saving grace of states and state transitions is that they are intentional; they are our way of thinking about what happens in nature. Symbol manipulation is a purely mental activity.

But Turing Machines and their equivalents offer an important insight. They identify symbol manipulation to be what we intuitively think of as computational activity. The Turing Machine model is our way of externalizing an entire class of mental activities, the class that we intuitively identify as symbolic.

In saying this we are separating (a) the sorts of *computational activities* characterized by Turing Machines, i.e., the Turing Machines themselves, from (b) the *class of functions* that these models compute, i.e., Turing computability. The various models of computational activities are all defined constructively, i.e., in terms of the operations one may perform *when constructing a computational procedure*. Furthermore, the equivalence proofs among the standard models are also constructive. We can constructively transform any Turing Machine into a recursive function and *vice versa*. The equivalence of these models shows that Turing Machines, recursive functions, etc. are equivalent *as programming languages*.

Computability theory then takes the generic class of software defined in this way and applies it to the task of computing functions. But this second step isn't necessary. What's important about the Church-Turing thesis is not the class of functions that can be computed but the possible programs one can write. Our revised version of the

Church-Turing thesis is that to be considered rigorous a thought process must, at least in principle, be expressible as a software.

## 4 Agent-based computing

The Turing Machine model is single threaded—as are the single processor von Neumann computers that we built based on it. But many of our computer science (and other) thought models are either parallel, asynchronous, or non-deterministic. Not all rigorously defined models are linear and single threaded. We have as yet been unable to build thought tools to help us externalize these kinds of non-deterministic computational ideas. Attempts to perform non-deterministic computations on a single-threaded computer result in unrealizable demands for resources.[11]

Four decades ago agent-based computing began to emerge Dahl [10]. Agent-based computing is an attractive form of asynchronicity because it relies on manageable parallelism—asynchronous computing threads that don't result in an unrealizable demand for computing resources. Its price is *chaotic asynchronicity*: minimally different event orderings may yield different results.

### 4.1 Open and far-from-equilibrium computing

Goldin and Wegner [11] have defined what they called *persistent Turing Machines* (and elsewhere *interaction machines*). These are Turing Machines that perform their computations over an indefinite period—continually accepting input and producing output without ever completing what might be understood as a traditional computation. Results of computations performed after accepting one input may be retained (on the machine's "working tape") and are available when processing future inputs. Although Wegner's focus is not on agent-based computing, his model is essentially that: agents which interact with their environments and maintain information between interactions. From here on we use the term *agent* to refer generically to an object that embodies a program which controls how it interacts with its environment.

Goldin and Wegner claim that their "interactive finite computing agents are more expressive than Turing machines." There has been much debate about this claim. We believe that to pose computability questions about agents is to ask the wrong question. We believe that what Wegner and Goldin have done is implicitly to have taken the same stance that we took explicitly above, i.e., to distinguish between the programs one can write and the functions those programs can compute.

In making this implicit distinction Wegner and Goldin point out that one need not think of the program that a Turing Machine embodies in functional terms, i.e., as closed with respect to information flow. One can also think of a Turing Machine as open with respect to information flow. This parallels the distinction in physics between systems that are closed and open with respect to energy flows. Wegner has outlined this position most recently in [12]. Complex systems are famously far from equilibrium with respect to environmental energy flows. Wegner and Goldin's interaction machines (and agents in general) are similarly far from equilibrium with respect to information flows.

---
[11] If we get it to work on a useful scale quantum computing may be the first such thought tool.

What might one gain from being open to information flows? An illustrative example is Prisoner's Dilemma (PD). If one were to develop an optimized PD player for a one-shot PD exchange—since it's one shot, the system is closed—it will Defect. Playing against itself, it will gain 1 point on each side—using the usual scoring rules. If one were to develop an optimized PD player to engage in an iterative PD sequence—the system is open—it will Cooperate indefinitely (presumably by playing a variant of Tit-for-Tat), gaining 3 points on each side at each time. Thus the same problem (PD) yields a different solution depending on whether one's system is presumed to be open or closed with respect to information flows.

### 4.2 Agents and their environments

Computation involves the interaction of a process with its environment. In all cases with which we are familiar, the environment is modeled as a simply structured collection of symbols, e.g., a tape, a grid, etc. None of these models are adequate when compared to the real-world environment within which we actually find ourselves. We do not know how to model the multi-scalar face that nature presents to us—but almost certainly it won't be as a tape or a grid.

- In our actual environment new entities and new kinds of entities may come into existence. We are able to perceive and interact with them. We are aware of no formal environmental framework capable of representing such phenomena.
- We do not understand the ultimate set of primitives—if indeed there are any—upon which everything is built.

We have referred [4] to these problems as the difficulty of looking upwards and the difficulty of looking downwards respectively.

We are just beginning [4] to understand the nature of entities and of the multi-scalar environment within which they exist. That environment involves entities on multiple levels, but it also involves forces at only the most primitive level. All other interactions are epiphenomenal. This is not simply a layered hierarchy, although it has some layered hierarchy properties.

Given our lack of understanding about these issues it is not surprising that we have not been able to develop a formal model of such an environment. Thus a fundamental open problem in computing is to develop a model of an environment that has the same sorts of multi-scalar properties as our real-life environment.

Our revised version of the Church-Turing thesis gives us confidence that our current understanding of agents as entities that embody programs is reasonably close to how we think about thinking. We are still quite far from the goal of formalizing appropriate environments within which to situate such agents.

### 4.2 The inevitable evolution and acceleration of intelligence

As we saw in the PD example, thinking in terms of open computation models leads to different results from thinking in terms of closed models. Yet both use the same class of possible programs—whatever is programmable in a general purpose programming language. Since open computation models include the class of Oracle machines, computability doesn't seem like the appropriate perspective when analyzing these systems. Is there another approach? We suggest that the notion of results achieved is more relevant. In the PD case, the result achieved is the number of points scored.

Under what other circumstances would it make sense to think of an agent in terms of results achieved? In [4] we discuss the nature of emergent entities. Static entities persist at an energy equilibrium in energy wells; but the more interesting dynamic entities persist only so long as they can extract energy from their environment.

Unfortunately most agent-based computer models either ignore the issue of energy or treat it very superficially. We believe that an integrated theory of energy and information would clarify how information flows enable evolution. A real-world agent would be a dynamic entity that embodied some software. If, through a random mutation, such an entity developed an enhanced ability to extract information from its environment then it will be more likely to survive—and, if capable of reproducing, reproduce. What evolves in this model is an enhanced ability to extract information from the environment. The need of dynamic entities for energy drives evolution toward increasingly more powerful informational processing capabilities.[12]

In this picture, information is being extracted from the environment at two levels. Each individual extracts information from the environment, which it processes as a way to help it find energy. Very simple real-life examples are plant tropisms and bacterial tendencies to follow nutrient gradients. More interestingly, the evolutionary process itself extracts information from the environment, which it then encodes (in DNA) as the "program" which individual agents use to process information from their environment. Thus the real intelligence is in the program, and the real information extracting activity is the evolutionary process which constructs the program.[13]

Can evolution itself evolve? Is there something that will enable an entity to extract information from the environment more effectively? Modern society stores information about how to process information from the environment as science. Can we go beyond science? Can the scientific process itself evolve? Science is the process of constructing mechanisms to extract and process information from the environment. Since science is a thought process, tools that enable us to externalize and improve our scientific thought processes will enhance our ability to do science.

---

[12] This seems to answer the question of whether evolution will always produce intelligence. It will whenever increased intelligence yields enhanced access to energy.

[13] Attempts to model this process have failed because their environments are too poor.

## 5  Conclusion

An environmentally sophisticated agent-based paradigm involves agents, each of which has the computing capability of a Turing machine, situated in an environment that reveals itself reluctantly. Such an agent in a real-world environment is like an Oracle machine, with nature as the oracle. Combining agents with dynamic entities yields real-world agents, which (a) must extract energy from their environment to persist and (b) embody software capable of processing information flows from the environment. The agent-based thesis is that this paradigm represents how, at the start of the 21$^{st}$ century, we have externalized our thoughts about our place with the world.

**Acknowledgment.** Many of the ideas in this paper were elaborated in discussions with Debora Shuger.